# Designing for Human Rights in AI

Evgeni Aizenberg[1,3] and Jeroen van den Hoven[2,3]

[1]Department of Intelligent Systems, Delft University of Technology, The Netherlands
E.Aizenberg@tudelft.nl

[2]Department of Values, Technology and Innovation, Delft University of Technology, The Netherlands

[3]AiTech multidisciplinary program on Meaningful Human Control over Autonomous Intelligent Systems, Delft University of Technology, The Netherlands

**ABSTRACT**

In the age of big data, companies and governments are increasingly using algorithms to inform hiring decisions, employee management, policing, credit scoring, insurance pricing, and many more aspects of our lives. AI systems can help us make evidence-driven, efficient decisions, but can also confront us with unjustified, discriminatory decisions wrongly assumed to be accurate because they are made automatically and quantitatively. It is becoming evident that these technological developments are consequential to people's fundamental human rights. Despite increasing attention to these urgent challenges in recent years, technical solutions to these complex socio-ethical problems are often developed without empirical study of societal context and the critical input of societal stakeholders who are impacted by the technology. On the other hand, calls for more ethically- and socially-aware AI often fail to provide answers for how to proceed beyond stressing the importance of transparency, explainability, and fairness. Bridging these socio-technical gaps and the deep divide between abstract value language and design requirements is essential to facilitate nuanced, context-dependent design choices that will support moral and social values. In this paper, we bridge this divide through the framework of Design for Values, drawing on methodologies of Value Sensitive Design and Participatory Design to present a roadmap for proactively engaging societal stakeholders to translate fundamental human rights into context-dependent design requirements through a structured, inclusive, and transparent process.

# 1. INTRODUCTION

Predictive, classifying, and profiling algorithms of a wide range of complexity — from decision trees to deep neural networks — are increasingly impacting our lives as individuals and societies. Companies and governments seize on the promise of artificial intelligence (AI) to provide data-driven, efficient, automatic decisions in domains as diverse as human resources, policing, credit scoring, insurance pricing, healthcare, and many more. However, in recent years the use of algorithms has raised major socio-ethical challenges, such as discrimination (Angwin et al., 2016; Barocas and Selbst, 2016), unjustified action (Citron and Pasquale, 2014; Turque, 2012), privacy infringement (Kosinski et al., 2013; Kosinski and Wang, 2018), spread of disinformation (Cadwalladr and Graham-Harrison, 2018), job market effects of automation (European Group on Ethics in Science and New Technologies, 2018a; Servoz, 2019), and safety issues (Eykholt et al., 2018; Wakabayashi, 2018).

It is becoming evident that these technological developments are consequential to fundamental human rights (Gerards, 2019; Raso et al., 2018) and the moral and social values they embody (e.g. human dignity, freedom, equality). The past decade has seen increasing attention to these urgent challenges within the technical, philosophical, ethical, legal, and social disciplines. However, collaboration across disciplines remains nascent. Engineering solutions to complex socio-ethical problems, such as discrimination, are often developed without a nuanced empirical study of societal context surrounding the technology, including the needs and values of affected stakeholders. As Selbst et al. (2019: 59) point out, most current approaches "consider the machine learning model, the inputs, and the outputs, and abstract away any context that surrounds [the] system." This gap puts technical implementations at risk of falling into the formalism trap, failing to "account for the full meaning of social concepts such as fairness, which can be procedural, contextual, and contestable, and cannot be resolved through mathematical formalisms" (Selbst et al., 2019: 61). On the other hand, calls for more ethically- and socially-aware AI often fail to provide answers for how to proceed beyond stressing the importance of transparency, explainability, and fairness. Bridging these socio-technical gaps is essential for designing algorithms and AI that address stakeholder needs consistent with human rights. We support the call of Selbst et al. to do so by shifting away from a solutions-oriented approach to a process-oriented approach that "draws the boundary of abstraction to include social actors, institutions, and interactions" (Selbst et al., 2019: 60).

In this paper, we bridge this socio-technical divide through the framework of Design for Values (Van den Hoven et al., 2015), drawing on methodologies of Value Sensitive Design (Friedman, 1997; Friedman et al., 2002; Friedman and Hendry, 2019) and Participatory Design (Schuler and Namioka, 1993). We present a roadmap for proactively engaging societal stakeholders to translate fundamental human rights into context-dependent design requirements through a structured, inclusive, and transparent process. Our approach shares the vision of Tubella et al. (2019) of applying Design for Values in AI and echoes other recent calls for incorporating societal and stakeholder values in AI design (Franzke, 2016; Rahwan, 2018; Simons, 2019; Umbrello and De Bellis, 2018; Zhu et al., 2018). In this work, we make an explicit choice of grounding the design process in the values of human dignity, freedom, equality, and solidarity, inspired by the Charter of Fundamental Rights of the European Union (EU) (*Official Journal of the European Union*, 2012). Our intention is to demonstrate how AI can be designed for values that are core to societies within the EU. Although human rights are not unproblematic or uncontroversial (Hopgood, 2013), the EU Charter represents the most widely shared and institutionalized consensual set of values and ethical principles among EU member states. It is worth noting that these values are not exclusive to EU member states and are shared by societies outside of the EU. At the same time, it is important to recognize at the outset that alternative value choices are likely in other cultures and socio-political systems. We should also make clear the distinction between legal interpretation of human rights through, for example, case law versus the notion of human rights as embodying particular moral and social values, such as human dignity. In this work we focus on the latter — human rights as moral and social values to be designed for. This distinction is significant because engaging stakeholders to design for values (rather than established law and standards) can lead to design requirements that go beyond the present scope of the law.

The presented design roadmap, with its anchoring in human rights, stakeholder involvement, and an established body of socio-technical design methodologies, is a tractable manifestation of the types of design approaches called for by the ethics guidelines of the European Commission's High-Level Expert Group on AI (2019) and the European Group on Ethics in Science and New Technologies (2018b). The following sections are organized as follows: Section 2 provides a brief introduction to Design for Values with an overview of stakeholder engagement methods from Value Sensitive Design and Participatory Design; in Section 3, we present our contribution, adopting fundamental human rights as top-level requirements that will guide the design process described in Section 2 and demonstrate the implications for a range of AI application contexts and key stakeholder considerations; in Section 4, we discuss future steps needed to implement our roadmap in practice.

## 2. DESIGN FOR VALUES

The term Design for Values (Van den Hoven et al., 2015) refers to the explicit translation of moral and social values into context-dependent design requirements. It encompasses under one umbrella term a number of pioneering methodologies, such as Value Sensitive Design (Friedman, 1997; Friedman et al., 2002; Friedman and Hendry, 2019), Values in Design (Flanagan et al., 2008; Nissenbaum, 2005), and Participatory Design (Schuler and Namioka, 1993). Over the past decades, these approaches have opened up the design focus from technological artifacts and technical requirements to include a proactive consideration of the societal context in which the technology is embedded, and how societal needs and values can be translated into socio-technical design requirements. Bringing the societal context into play is often characterized by empirical research and proactive involvement of direct and indirect stakeholders affected by the technology throughout the design process. We would like to highlight that design requirements can be both social and technical in nature (hence socio-technical), reflecting an interplay of human and technical artifacts and processes. That includes interactions between humans and technical artifacts, but also importantly human-to-human interactions and organizational practices.

The gradual translation of abstract values into design requirements is referred to as value specification and can be visually mapped as a values hierarchy (Van de Poel, 2013) (Figure 1a). First, values are expanded into norms, which are properties or capabilities that the designed technology should exhibit in order to support desired values. For instance, the value privacy (Figure 1b) may in a certain context be interpreted as implying informed consent, confidentiality, and right to erasure. Thus, these three norms are viewed by the stakeholders as the specification of what the value privacy means in that context of use. Each of these norms is in turn specified further into a socio-technical design requirement. For example, in Figure 1b, informed consent to processing of personal data is to be implemented as positive opt-in (requires explicit permission) and confidentiality is to be implemented in terms of homomorphic encryption techniques. These again can be further specified.

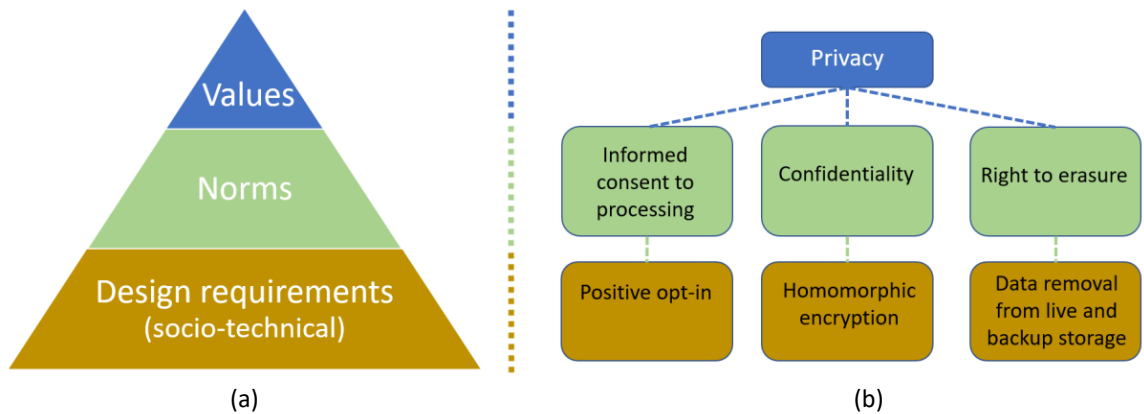

**Figure 1.** Values hierarchy (a) visually maps the context-dependent specification of values into norms and socio-technical design requirements. For instance, the value privacy (b) may in a certain context be interpreted by stakeholders as the ability of individuals to provide informed consent to processing of their personal data, its confidentiality, and the ability to erase personal data from the system. In turn, each of these properties is linked to a design requirement to be implemented.

Note that the relation between higher levels and lower levels in the hierarchy is not deductive (Van de Poel, 2013). The norms in the example above may be insufficient or entirely irrelevant interpretations of privacy in a different context of use. Parties to the debate about privacy may disagree with the proposed decomposition. The value hierarchy offers a structure to have a focused debate about these disagreements. Detailed proposals for change can be made and discussed between stakeholders. Therefore, value specification provides an explication of an abstract value and is always context-dependent. In the opposite direction, the link between a lower level element ($L$) in the hierarchy and a higher level element ($H$) is characterized by a *for the sake of* relation. $L$ being done *for the sake of H* can embody a number of more specific relations (Van de Poel, 2013):

1) $L$ is a means to $H$.
2) $L$ is a goal, the achievement of which contributes to the achievement of $H$.

3) *L* enables the achievement of *H* without itself contributing directly to that achievement.
4) *L* removes an obstacle to *H*.

Each of these relations can be further elaborated by the stakeholders in a specific context. For example, what are the societal, ethical, and human rights considerations that lead stakeholders to conclude that a right to erasure is a means to privacy? Hence, the *for the sake of* relation captures stakeholders' contextual reasoning and motivations for requiring that a higher level element *H* (e.g. privacy) is supported by a lower level element *L* (e.g. right to erasure). This nuanced, context-dependent, stakeholder-formulated argumentation of design requirements is an essential mechanism for bridging the socio-technical gap. It facilitates an inclusive, socially-aware, cross-disciplinary conversation necessary to account for the breadth and depth of complex socio-ethical issues and value implications.

A values hierarchy exposes in a structured and transparent fashion which moral and social values (normative requirements) the technology in question should support, what are the stakeholder interpretations of those values in the specific context of use, and what are the corresponding design requirements to be implemented. This allows all stakeholders — individuals affected by the technology's operation, direct users, engineers, field experts, legal practitioners, etc. — to debate design choices in a manner that traces the reasons, advantages, and shortcomings of each choice to societal norms. Because of the traction and perspicuous representation of moral reasons that such a framework can provide, moral discussions about engineering design can contribute to moral learning, technical improvement, social legitimacy, and trust building between parties. In the following subsections, we review specific methods from Value Sensitive Design (Friedman and Hendry, 2019) and Participatory Design (Schuler and Namioka, 1993) that engage stakeholders to iteratively specify the requirements in the different levels of the hierarchy.

## 2.1 Tripartite methodology

The structure of the design process we propose to follow is based on the tripartite methodology of Value Sensitive Design, which integrates three types of investigations (Friedman et al., 2013):

> *Conceptual investigations* identify the stakeholders and values implicated by the technological artifact in the considered context and define value specifications.

*Empirical investigations* explore stakeholders' needs, views, and experiences in relation to the technology and the values it implicates.

*Technical investigations* implement and evaluate technical solutions that support the values and norms elicited from the conceptual and empirical investigations.

It is important to highlight that the three types of investigations are iterative and integrative (Friedman et al., 2013) in that they are "meant to inform each other rather than be engaged as separate, modular activities. Investigations may overlap, happen in different orders, or intertwine with each other" (Davis and Nathan, 2015: 16). For example, a follow-up empirical investigation after an initial prototype implemented in the technical investigation may reveal value tensions that were not identified earlier, resulting in a new iteration of conceptual investigation.

## 2.2 Engaging with stakeholders

Stakeholders are engaged through a range of empirical methods, such as surveys, interviews, participant observation (Musante (DeWalt), 2015), card methods (Friedman and Hendry, 2012; Wölfel and Merritt, 2013), and participatory prototyping (Lim et al., 2008), to elicit their "understandings, concerns, reflections, and aspirations" (Davis and Nathan, 2015: 16). One useful instrument that can be combined with interviews is value scenarios (Nathan et al., 2007), which are nuanced imagined scenarios involving the proposed technology and various direct and indirect stakeholders. Value scenarios can help uncover "nefarious and unusual uses, value tensions, and longer-term societal implications that might otherwise go unnoticed" (Czeskis et al., 2010: 3). During interaction with stakeholders, Participatory Design encourages co-designers to explore alternative visions and be sensitive to silences (Van der Velden and Mörtberg, 2015). Exploring alternative visions involves postponing what may seem as the obvious technical solution to a complex socio-ethical problem in order to enable stakeholders "who otherwise might be invisible or marginalized to have a voice in the design process" (Van der Velden and Mörtberg, 2015: 53). This is yet another means through which the socio-technical gaps are bridged. It is also possible that during the design process some stakeholders may turn silent or be silenced. Therefore, it is essential to respect and be sensitive to silences, recognizing them as cues for taking a step back and reaffirming an inclusive, "democratic design space, which gives voice to all" (Van der Velden and Mörtberg, 2015: 54).

## 2.3   Local meaning and context

Various authors have highlighted the moral importance of accommodating local meanings of social practices and language, e.g. (MacIntyre, 1981; Wittgenstein, 1953). Actions and language have meaning in a particular social practice, or institutional setting, which has a purpose or goal and has its own norms governing behavior in that setting. Our representations and design need to respect these social and normative contexts. In that regard, a key question that will often emerge during Design for Values in AI is to what extent an AI system should replace the actions of a human agent in performing a certain activity, considering the social, normative, and institutional purpose of that activity. A technique that can help address this question in a structured manner is what has been called "the Nature-of-Activities" approach (Santoni de Sio et al., 2014), which divides an activity into its *goal-directed* and *practice-oriented* aspects. To illustrate this, consider the example of using care robots for the activity of lifting a patient from a hospital bed, studied by Santoni de Sio and Van Wynsberghe (2016). This activity has the *goal-directed* aspect of safely raising the patient out of bed at a certain angle and speed and safely placing her/him in a wheelchair. At the same time, the performance of this activity by a human care giver has the *practice-oriented* aspect of developing the bond and relationship between the care giver and the patient. This relationship has a social value for the patient and is also important for the patient's long-term care: "to be honest about their symptoms, to take their medication and to comply with their care plan" (Santoni de Sio and Van Wynsberghe, 2016: 1752).

As another example, consider a university that would like to automate the task of advising students' curriculum choices with algorithm-generated personalized curriculum plans. Normally, this activity is performed by a trained, experienced academic counselor in a one-on-one session with the student. It has the *goal-directed* aspect of recommending a curriculum that fits the students' choice of specialization, past courses, and grades. But universities are more than efficient information transmission institutions. Their practices and life forms, their telos and supreme value is the pursuit of knowledge by building communities of learning and scholarship, which require open communication and trust. The advising activity has therefore also the *practice-oriented* aspect of building an informal and socially interactive companionship between the student and a human counselor, who can help the student elicit interests that are outside of her/his initial comfort zone and provide emotional support in moments of anxiety.

Adopting this type of perspective in the study of context helps to bridge socio-technical gaps by revealing valuable human-to-human interactions that may otherwise go unnoticed. Doing so can help identify which activities (or sub-tasks of an activity) are best delegated to an AI system and which are best delegated to a human. In fact, innovative interactions between humans and AI can emerge, where the division of tasks between the two satisfies stakeholder needs and values to a greater extent compared to either the human or AI performing the activity alone.

## 3. HUMAN RIGHTS AS TOP-LEVEL REQUIREMENTS

Our roadmap of Designing for Human Rights in AI begins with adopting the four values at the basis of the EU Charter of Fundamental Rights (*Official Journal of the European Union*, 2012) — human dignity, freedom, equality, and solidarity — as top-level requirements that will guide the design process described in the previous section. We demonstrate the implications of values embodied by the Charter's provisions for a range of AI application contexts and key stakeholder considerations, such as autonomy over self-representation, non-discrimination, privacy, transparency, and job market effects of automation. When viewed through the perspective of a values hierarchy, what follows are the elements of the top level of the hierarchy and an analysis of how values embodied by the Charter's human rights provisions relate to and support each other.

### 3.1   Dignity

Human dignity is a foundational value in the EU Charter and is the central, overarching human value at stake in AI, be it in the context of discrimination, unjustified action, privacy violations, or job market transformations. As stated in the Charter's official guiding explanations (*Official Journal of the European Union*, 2007: C 303/17): "dignity of the human person is not only a fundamental right in itself but constitutes the real basis of fundamental rights." In this sense, dignity is a highly multifaceted concept. Ironically, one way to grasp the broad range of its meanings is to explore what humans consider to be violations of their dignity, as done by Halbertal (2015), who points out the following three categories of violations:

1) *Humiliation*: being put in a state of helplessness, insignificance; losing autonomy over your own representation.
2) *Instrumentalization*: treating an individual as exchangeable and merely a means to an end.
3) *Rejection of one's gift*: making an individual superfluous, unacknowledging one's contribution, aspiration, and potential.

The AI-related impact on dignity can be more clearly seen through these three dimensions. For example, discrimination and unjustified action lead to humiliation, putting individuals in a state of helplessness when faced with opaque and biased algorithmic decisions. Job market effects of automation may cast whole communities of workers as superfluous and exchangeable for the more profitable alternative of AI. Large-scale collection of data for online advertisements exploits individuals as exchangeable data points that are merely a means to driving profits. As argued by the European Data Protection Supervisor (2015: 12), "better respect for, and the safeguarding of, human dignity could be the counterweight to the pervasive surveillance and asymmetry of power which now confronts the individual. It should be at the heart of a new digital ethics."

By examining dignity along the dimensions through which it can be violated, the human rights provisions of the EU Charter can be seen as providing a collective and mutually complementing response to the question of how to protect against violations of human dignity. In other words, the Charter is a manifestation of both rights and limitations which are in place *for the sake of* upholding human dignity. Table 1 highlights some of the Charter provisions that are especially relevant in AI contexts. To reflect its central, overarching role, dignity is placed at the root of our design roadmap as a first-order value (Figure 2). On the second order, follow four key values fulfilling a *for the sake of* relation with respect to dignity: freedom, equality, solidarity, and right to life. Note that here we extend the use of the *for the sake of* relation to convey the intra-level relationships between different values. Placing the right to life (Article 2 within the Dignity Title) on the same order as freedom, equality, and solidarity is motivated by its direct relation to safety, which is a crucial consideration in AI contexts. The other three values have their own dedicated Title sections within the Charter, whose implications for AI we explore below.

**Table 1.** Highlighted provisions from the Dignity, Freedom, Equality, and Solidarity Titles of the EU Charter (*Official Journal of the European Union*, 2012).

| | |
|---|---|
| **Dignity** (Title I) | (*Article 1, Human dignity*): Human dignity is inviolable. It must be respected and protected.<br><br>(*Article 2, Right to life*): Everyone has the right to life. |
| **Freedom** (Title II) | (*Article 6, Right to liberty and security*): Everyone has the right to liberty and security of person.<br><br>(*Article 8, Protection of personal data*): Everyone has the right to the protection of personal data concerning him or her. Such data must be processed fairly for specified purposes and on the basis of the consent of the person concerned or some other legitimate basis laid down by law. Everyone has the right of access to data which has been collected concerning him or her, and the right to have it rectified.<br><br>(*Article 11, Freedom of expression and information*): Everyone has the right to freedom of expression. This right shall include freedom to hold opinions and to receive and impart information and ideas without interference by public authority and regardless of frontiers. |
| **Equality** (Title III) | (*Article 21, Non-discrimination*): Any discrimination based on any ground such as sex, race, color, ethnic or social origin, genetic features, language, religion or belief, political or any other opinion, membership of a national minority, property, birth, disability, age, or sexual orientation shall be prohibited. |
| **Solidarity** (Title IV) | (*Article 34, Social security and social assistance*): The Union recognizes and respects the entitlement to social security benefits and social services providing protection in cases such as maternity, illness, industrial accidents, dependency or old age, and in the case of loss of employment, in accordance with the rules laid down by Union law and national laws and practices…<br>In order to combat social exclusion and poverty, the Union recognizes and respects the right to social and housing assistance so as to ensure a decent existence for all those who lack sufficient resources, in accordance with the rules laid down by Union law and national laws and practices. |

## 3.2 Freedom

The content of the Freedom Title is strongly associated with individual autonomy. For example, the right to liberty in Article 6 protects individuals against arbitrary deprivation of physical freedom (*Official Journal of the European Union*, 2007). In the context of algorithmic systems, one sensitive area that is directly implicated by Article 6 is criminal justice, e.g. predictive policing systems and recidivism risk assessments. Predictions and judgements made by these algorithms are typically based on statistical correlations rather than causal evidence. This means that a person may be labeled as high risk simply because her/his demographics correlate with data gathered on previously arrested people. This clash of correlation vs. causation (Kitchin, 2014; Rieder and Simon, 2017) brings out the wider issue of data determinism, a situation in which data collected about individuals and the statistical inferences drawn from that data are used to judge people on the basis of what the *algorithm says* they *might do/be/become*, rather than what *the person* actually *has done* and *intends to do* (Broeders et al., 2017). Data determinism therefore contributes to statistical dehumanization (Van den Hoven and Manders-Huits, 2008) and is directly at odds with the autonomy of individuals to define themselves and their aspirations — autonomy over their own representation. Existing examples of such inferences go well beyond criminal justice and include talent and personality assessments for job candidate selection, employee performance assessments, financial credit scoring, and many more areas. Understanding the effects of these systems on autonomy over self-representation requires a critical look together with the affected stakeholders at the contextual nature of what an algorithm attempts to predict or evaluate — the target variable (Barocas and Selbst, 2016) — and its causal links to data used on the input. The conceptions of target variables in the examples above (criminal risk, personality traits, job performance, credit risk) are highly fluid, contextual, and contestable (Selbst et al., 2019). Algorithms remove this fluidity, context, and contestability by attempting to fix the conception to a set of quantifiable inputs and outputs. As a result, it is the algorithm's conception that takes center stage, and that directly infringes people's autonomy over self-representation, leading to data determinism and discrimination (Citron and Pasquale, 2014; Leese, 2014; Mann and Matzner, 2019).

The illustrated impact of data determinism underlines that the automatic and quantitative nature of AI must not be confused with objectivity. This also brings out the interplay between contestability, justification, and freedom. If the output of an employee performance assessment algorithm is uncontestable, receiving a high score requires employees to adapt their behavior in a manner that would conform with the algorithm's conception of a 'good employee'. In other words, in order to achieve a desired outcome, the freedom to act true to one's own personality and wishes is compromised. As Wachter and Mittelstadt (2019: 20) argue, "such chilling effects linked to automated decision-making and profiling undermine self-determination and freedom of expression and thus warrant more control over the inferences that can be drawn about an individual." Furthermore, people's freedoms can be compromised without their awareness of being subjected to algorithmic profiling (Hildebrandt, 2008). As Hildebrandt (2008: 318) elaborates, "(1) an abundance of correlatable data and (2) the availability of relatively cheap technologies to construct personalised knowledge out of the data, create new possibilities to manipulate people into behaviour without providing adequate feedback on how their data have been used." Therefore, individuals' ability to enforce their autonomy significantly relies on their awareness of being subjected to algorithmic profiling and their ability to contest the rationale behind algorithmic decisions. In turn, for individuals to contest an algorithm and ensure that the burden of proof is first of all placed on the algorithm's operator (Wachter and Mittelstadt, 2019), socio-technical systems need to be sufficiently transparent to provide context-dependent legible justifications for their decisions and allow for case-specific discretion. In order to design for such transparency, it is essential to shape an understanding of what constitutes a legible justification and effective contestation mechanism together with societal stakeholders who would perform the contesting task in a given context.

The second highlighted section from the Freedom Title, Article 8, provides important data privacy protections. The Article explicitly states that collected data should be processed "for specified purposes and on the basis of the consent of the person concerned or some other legitimate basis laid down by law" (*Official Journal of the European Union*, 2012: C 326/397). Additionally, every individual is provided with the right of "access to data which has been collected concerning him or her" (*Official Journal of the European Union*, 2012: C 326/397). In AI contexts, these provisions have profound implications for privacy and transparency, considering the growing pervasiveness of profiling algorithms that base their outputs on large-scale data collection (Citron and Pasquale, 2014). There is again a substantive link here to individuals' autonomy over self-representation, since privacy is essential for exercising that autonomy and protection from shaming or distortion of one's personality (Velleman, 2005). This is something also understood as informational self-determination (Cavoukian, 2008). Designing for these provisions requires stakeholder consensus around which data are justifiably necessary for performance of a processing task, and which mechanisms should be in place for individuals to exercise access and conscious consent to the use of their personal data.

We finalize the analysis of the Freedom Title with Article 11, which focuses on freedom of expression and information, ensuring that individuals have the autonomy of sharing and obtaining information without arbitrary impediment. On the information sharing front, this provision is directly linked to the ongoing debate on balancing freedom of speech versus hate speech, public shaming, and disinformation on social media and the Internet at large. On the information receiving front, these issues are intertwined with content personalization algorithms, such as news recommender systems, and their impact on information diversity (Bay, 2018; Pariser, 2011), a necessary condition for a functioning democracy. These are examples of value tensions between different stakeholders. Explicitly recognizing these types of tensions and engaging in nuanced, context-specific deliberations that seek to resolve them are integral to the design process.

### 3.3 Equality

At the heart of the Equality Title is the principle of non-discrimination in Article 21, which explicitly prohibits any discrimination on grounds such as ethnical origin, gender, skin color, religion, sexual orientation, political opinions, disability, and more. For AI, this is another crucial consideration. Given the scale and speed at which these systems operate, algorithms that systematically disadvantage one demographic group or community vs. another can exacerbate historical discrimination, socio-economic gaps, and inequity (Citron and Pasquale, 2014; O'Neil, 2016). Designing non-discriminating algorithms is particularly challenging considering that discrimination can occur even when protected personal attributes, as those listed in the Equality

Title, are excluded from the data (Barocas and Selbst, 2016) and the fact that non-discrimination is a highly contextual, contestable, and procedural concept that does not always lend itself to mathematical formalisms (Selbst et al., 2019). Therefore, designing for equality entails developing a nuanced understanding of present and historical social dynamics that contribute to discrimination in a given context and assessment of both social (e.g. organizational processes, policies, human-to-human interaction) and socio-technical means to address it. On the technical side, special attention needs to be devoted to discriminatory implications of choices for target variables, class labels, data collection gaps across different communities, and feature selection (Barocas and Selbst, 2016). Furthermore, algorithmic profiling can lead to new types of discrimination (Leese, 2014; Mann and Matzner, 2019) that systematically disadvantage individuals based on input data representations that have no direct connection to familiar protected grounds, such as gender or ethnical origin — consider the discussed examples of talent and employee performance assessments. Finally, designing for equality also entails designing for privacy to avoid situations in which exposure of personal data results in unjustified preferential treatment (Van den Hoven, 1997). This aspect once again requires stakeholders to reflect on what data are justifiably necessary for the performance of a given data processing task.

### 3.4 Solidarity

The Solidarity Title effectively stipulates a responsibility of members of society towards each other in "combat[ing] social exclusion…so as to ensure a decent existence for all those who lack sufficient resources" (*Official Journal of the European Union*, 2012: C 326/402). It recognizes that individuals' ability to exercise their rights, and therefore uphold their dignity, can be compromised in circumstances such as "maternity, illness, industrial accidents, dependency or old age, and in the case of loss of employment" (*Official Journal of the European Union*, 2012: C 326/402). This has important implications for design of AI. First, algorithmic systems that operate based on statistical correlations, by their nature make judgements based on the relative position of data collected about an individual with respect to the population sample the algorithm was trained on. In other words, what matters is not the substance of circumstances and experiences of that individual (e.g. loss of employment, single parenthood, disability), but rather how they compare to the population sample — Vedder (1999) refers to this as de-individualization. In many situations, this contradicts the empathetic spirit of the Solidarity Title because a nuanced, humane consideration of the dignity of the individual in question is replaced by an attempted mathematical measurement of where that individual stands compared to others. This again brings out the infringement on autonomy over self-representation by data determinism and the formalism trap issues discussed above. As a result, there is substantial risk that individuals and communities whose societal circumstances compromise their ability to exercise their rights may end up being pushed further to the fringes, introducing self-perpetuating feedback loops and increasing socio-economic disparity and inequity (Citron and Pasquale, 2014; O'Neil, 2016). Some vivid examples are algorithms used for insurance pricing, financial credit scoring, and predictive policing. As in the case of the value freedom, autonomy over self-representation and transparency of decision-making (including legible justification and effective contestation mechanisms) are essential in design for solidarity. Furthermore, to account for the far-reaching societal and economic impacts on various communities, relevant field experts should participate together with affected stakeholders in the translation of their concerns into context-dependent design requirements.

The second major implication of the Solidarity Title relates to the impact of AI systems on the job market. Integration of decision-support systems, assisting robots, and self-driving vehicles has the potential to shift many workers out of their jobs, causing anxiety and political unrest. Involvement of direct and indirect stakeholders is essential to uncover both short and long-term implications of replacing entire jobs or automating specific tasks within a job. As illustrated in Section 2.3, it can be helpful to construct a nuanced understanding of the harm-benefit tradeoffs of automation, revealing important stakeholder considerations that are not obvious at first sight. These tradeoffs may lead to a realization that designing for solidarity will involve interactions between humans and AI where the machine does not replace the human, but rather assists the human worker to perform her/his tasks in a more efficient and safer manner, perhaps even enhancing valuable human-to-human interactions. Furthermore, in some contexts the tradeoffs will be dynamic in time. For example, consider a scenario in which public transport companies are deciding on the timeline for replacing their fleet with self-driving vehicles. Although a fast replacement may be feasible and profitable to the companies, it will result in large-scale layoffs, with many workers unlikely to be able to retrain to other available jobs on the market. On the other hand, a gradual replacement of the fleet would allow more transition time, during which employers and workers can develop training programs that will help workers extend their skillset and knowledge to be competitive in the changing job market. This example of course does not do justice to the full range and complexity of considerations at stake, but it illustrates the necessity to engage with needs and value tensions between different stakeholders in a nuanced and inclusive manner.

### 3.5 Third-order values and context-dependent specification

As illustrated by the above implications of fundamental human rights for AI systems, the second-order values in our design roadmap can be specified further into third-order values that are more familiar in discussions surrounding the ethics of AI: non-discrimination, privacy, transparency, and safety (Figure 2). This list is not exhaustive and is open to expansion and refinement. Notice that some of these third-order values fulfill a *for the sake of* relation with respect to more than one second-order value, reflecting the value relationships analyzed above. For example, the value privacy supports both freedom and equality, while transparency supports both freedom and solidarity. From here onward, further specification of these values and their translation into system properties becomes highly context dependent. This is precisely where the methods outlined in Section 2 with the crucial involvement of societal stakeholders enter to elicit the normative and socio-technical implications of the higher order values in a specific context. As noted earlier, this design process requires iterating through several rounds of conceptual, empirical, and technical investigations. With each iteration, a more nuanced and broad understanding of normative requirements, value tensions, and socio-technical design requirements will emerge. The iterative, adaptive, and inclusive nature of this process is analogous to key principles behind Agile software development (Beck et al., 2001). This offers perhaps an unexpected opportunity to recognize similarities and possible bridges between practices employed by software engineers and the design approach presented here.

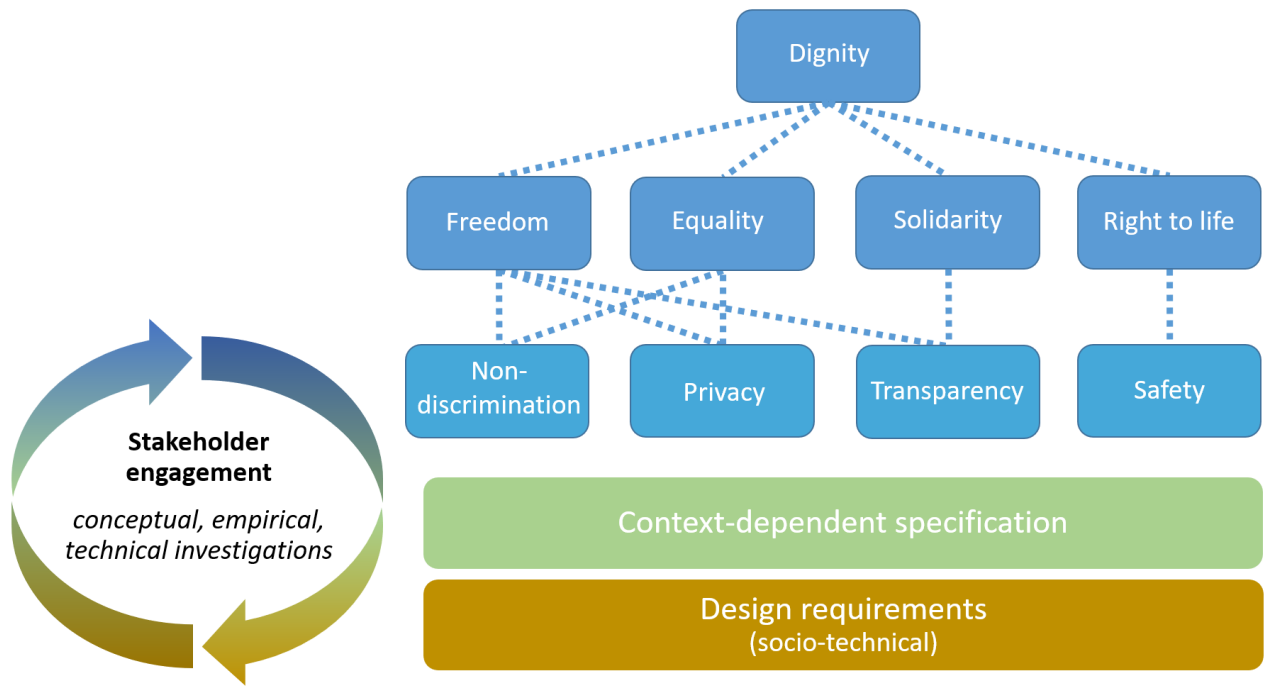

**Figure 2.** Human rights as top-level requirements in the design roadmap. The dashed lines depict *for the sake of* relations between individual values. Further specification of values into context-dependent design requirements is achieved through stakeholder engagement, iterating through conceptual, empirical, and technical investigations.

It is to be expected that in some contexts and case scenarios the stakeholders may reach the conclusion that none of the proposed technical implementations sufficiently satisfy important normative requirements. On the one hand, this can be an opportunity for technical innovation, motivating outside of the box thinking that can lead to a successful resolution (Van den Hoven et al., 2012). On the other hand, as Selbst et al. (2019) caution, we should not fall into the solutionism trap (Morozov, 2013), assuming that some form of AI is always the solution to the problem or goal at hand. In fact, in situations where normative requirements cannot be matched with suitable implementations, the hierarchical nature of our design roadmap provides the means to trace exactly which norms, values, and ultimately human rights would be insufficiently supported or violated. In such scenarios, designing responsibly with respect to individuals' human rights will demand recognizing that introduction of AI in that context will be harmful. The notion of rejecting AI solutions which fail to sufficiently fulfill normative requirements would not be different from the manner in which other technologies, such as cars, airplanes, home appliances, and bridges are required to fulfill various certification standards whose purpose is to support values such as safety, accessibility, and environmental protection.

## 4. FROM ROADMAP TO PRACTICE, MOVING FORWARD

The roadmap presented in this paper opens up a range of practical questions concerning its implementation. For example, how and who determines the appropriate stakeholders in a given context? How to assess consensus among stakeholders? How and who determines if sufficient iterations have been made throughout the conceptual, empirical, and technical investigations? Is there a need for an external body or institution that would certify the integrity of the design process and ensure its transparency with respect to society?

Some of these questions can be addressed by knowledge and experience accrued from past applications of Design for Values in various domains (Azenkot et al., 2011; Czeskis et al., 2010; Friedman et al., 2006; Woelfer et al., 2011). However, many of these questions bring us to uncharted waters, much like AI itself. Moving forward depends on our willingness to put ambitious multidisciplinary ideas to the test, learn to communicate efficiently with stakeholders of diverse backgrounds, and build together the necessary socio-technical checks and balances at different scales. The starting point would be case studies focused on concrete contexts and stakeholders. Valuable knowledge can then be gathered about the performance of empirical methods for stakeholder engagement, the context-dependent requirements that emerge, and the socio-technical solutions and human-AI interactions the design roadmap leads to. With this interdisciplinary knowledge and experience we can transition from case studies to specific design protocols and certification procedures applicable to different social domains. As concluded in a recent Council of Europe report on discrimination and AI (Zuiderveen Borgesius, 2018: 39), there is a need for "sector-specific rules, because different values are at stake, and different problems arise, in different sectors." In the long term, applying the human rights-based design roadmap as a basic building block in different contexts can gradually assemble the complex mosaic of AI that functions in synergy with fundamental human rights.

There is also significant overlap in the spirit and substance of the presented design approach and the human rights due diligence and impact assessment processes advanced by the United Nations Guiding Principles (UNGP) on Business and Human Rights (United Nations, 2011). Both seek to proactively engage relevant stakeholders (UNGP Principle 18) in a nuanced debate about the societal impact of the technology in question and resolve identified risks (UNGP Principle 19). In the context of AI, there is a particularly strong connection between our approach and the Human Rights, Ethical and Social Impact Assessment (HRESIA) presented by Mantelero (2018) and the Algorithmic Impact Assessment presented by Reisman et al. (2018). HRESIA also puts emphasis on the socio-ethical dimension of human rights, going beyond a strictly legal assessment of implications, and stresses the importance of context study, including consultation with potentially affected stakeholders. It is important to highlight that while identifying negative impact (to put simply, what not) is certainly an indispensable part of Design for Values, at the heart of it is translating stakeholders' positive needs and values (to put simply, what yes) into concrete design choices that support these needs consistent to values. Therefore, we believe that our design approach can complement human rights due diligence processes by providing a structured and inclusive path for formulating design requirements that address the concerns, needs, and expectations identified in a human rights impact assessment or an algorithmic impact assessment. For example, the values hierarchy mapping can help facilitate the debate among stakeholders by linking concrete components of the system in question with normative requirements they support. Such a decomposition of a complex socio-technical system helps to bridge between different levels of abstraction and makes the conversation among stakeholders more tractable, legible, and transparent. Thus, with its anchoring in a body of established inclusive design methodologies, such as Value Sensitive Design and Participatory Design, our approach offers practical tools that can support and further advance human rights due diligence processes consistent to the UNGP.

Positioning the presented design approach within the framework of human rights due diligence can also help clarify the question of responsibility and roles of the state and private actors. According to the notion of corporate responsibility in the UNGP, business enterprises have a responsibility to avoid causing or contributing to adverse human rights impacts and address such impacts when they occur (UNGP Principle 13a). The presented design approach can help companies meet this responsibility. More concretely, it provides a process to proactively study the societal context, forming a nuanced understanding of stakeholder needs and human rights implications, and formulate design specifications in consultation with affected stakeholders. More broadly, we believe that designing for human rights is first and foremost the responsibility of the entity, private or public, that introduces an AI system within a given societal context. At the same time, just as with various types of impact assessments, the state and democratic institutions should play a crucial role in oversight, verification, and adjudication of the design process. This includes mechanisms for adjudicating unresolved disputes between stakeholders and ensuring transparency of the design process in a manner that allows for public scrutiny. This is consistent with the notion of state duty in the UNGP. Thus, positioning the presented design approach within the framework of human rights due diligence can help address a number of practical questions related to responsibility and adjudication posed at the beginning of this section. Moving forward, it is of great interest to explore this in practice.

## 5. CONCLUSION

Major socio-ethical challenges raised by AI systems in recent years, such as discrimination, unjustified action, and job market effects of automation, have made it clear that this technology has direct impact on people's fundamental human rights. It is for that reason that in this work we call for a proactive approach of Designing for Human Rights in AI. We have shown how the values at the basis of the EU Charter of Fundamental Rights — human dignity, freedom, equality, and solidarity — relate to major issues at stake and translate into normative requirements that can guide a structured, inclusive, and transparent design process that will involve diverse stakeholders and help bridge existing socio-technical gaps.

As noted at the outset, we made an explicit choice of focusing on values that are core to societies within the EU, as embodied by the EU Charter. Our view is that given the complex reality that interpretation of international human rights norms varies across different regions and socio-political systems, the ways in which different societies will design and incorporate AI technology will vary accordingly. Considering this diversity, had we based our analysis on a more global human rights instrument, such as the Universal Declaration of Human Rights, the analysis would likely have to be even more abstract to steer clear of imposing a particular geo-political perspective and call for follow-up studies on a more regional level. Therefore, while openly recognizing the limitations and inherent normative biases of the analysis presented in this paper, we focus on the more regionally coherent level of the EU. That said, it should be recognized that the EU itself comprises of diverse cultures and traditions. However, the very tangible presence of the EU as a socio-political, legal fabric that unites these diverse countries under the values embodied by the EU Charter, allows for a tangible analysis of AI implications and design considerations with respect to human rights the way they are perceived across the EU. At the same time, the values of human dignity, freedom, equality, and solidarity are also core to many societies outside of the EU. Therefore, many of the implications of these values on AI design discussed in this paper are applicable in other societies. To be clear, we have no intention of imposing the EU view on human rights onto other regional contexts. We sincerely hope that this work can serve as an example of an inclusive design process that can be applied in other regional contexts, leading to designs that support human rights.

Although this work is motivated in big part by the impact of AI on human rights, we should recognize that existing human institutions are highly imperfect themselves. However, it is also the case that despite these imperfections, significant progress has been made in advancement of human rights around the world during the 20th century and even over the past decades. And yet, there is a long way ahead. This again brings out the importance of designing from a joint socio-technical perspective. Technology can play a helpful role in advancing human rights, but as the discussed solutionism trap warns us, we should not mislead ourselves into believing that complex societal problems can be fixed by "smart" technology alone, which is often wrongly assumed to be ethically neutral and objective. It is therefore our hope that the presented design approach allows societies and institutions to engage in an inclusive and structured debate about how to advance the enjoyment of human rights in a manner that discovers the appropriate role AI can play in supporting humans. Designing for Human Rights should not be viewed as an obstacle to technical innovation. On the contrary, we believe that it is key to achieving innovative interactions between humans and AI systems in a manner that satisfies stakeholder needs consistent to moral and social values embodied by human rights. That does not mean that it is an easy path to tread: learning to communicate across disciplines, resolving value tensions between stakeholders, and recognizing that more automation and AI is not always the desired solution. However, the long-term benefits of this approach can be shared both by individuals whose dignity will be held paramount in a changing socio-technical landscape and engineers whose products will enjoy a higher level of acceptance and trust.

**ACKNOWLEDGEMENTS**

We would like to thank Inald Lagendijk for his critical perspective and helpful input in developing this paper. We also want to express our gratitude to the anonymous reviewers for their valuable feedback.